\documentclass[twocolumn]{aastex63}
\usepackage[]{units}
\usepackage[update,prepend]{epstopdf}
\usepackage{graphicx}
\usepackage{amssymb}
\usepackage{amsmath}
\usepackage{threeparttable}
\usepackage{hyperref}
\usepackage{color}

\usepackage{mathtools}

\usepackage{makecell}

\newcommand{\Msun}{\,M$_\odot$}

\mathchardef\mhyphen="2D
\shorttitle{Wide tertiaries around close binaries}
\shortauthors{Hwang}

\begin{document}

\title{
The mystery in Gaia DR3 triples: occurrence rates, orientations, and eccentricities of wide tertiaries around close binaries
}

\correspondingauthor{Hsiang-Chih Hwang}
\email{hchwang@ias.edu}
\author[0000-0003-4250-4437]{Hsiang-Chih Hwang}
\affiliation{School of Natural Sciences, Institute for Advanced Study, Princeton, 1 Einstein Drive, NJ 08540, USA}

\begin{abstract}

The formation of close binaries has been an open question for decades. A large fraction of close binaries are in triple systems, suggesting that their formation may be associated with the Kozai-Lidov mechanism. However, this picture remains under debate because the configurations of many observed triples are unlikely to trigger the Kozai-Lidov mechanism. In this paper, we use the close binary samples, including eclipsing, spectroscopic, and astrometric binaries, from Gaia Data Release 3 to investigate the mysterious connection between inner binaries and their wide tertiaries. We show that the wide tertiary (at $10^3$-$10^4$\,AU) fraction increases with decreasing orbital periods of the inner binaries. The wide tertiary fraction of eclipsing binaries (a median orbital period of $0.41$\,day) is $2.33\pm0.11$ times higher than the field wide binary fraction. Furthermore, there is a tentative excess at $\sim10^4$\,AU for tertiaries of eclipsing binaries. Using the $v$-$r$ angle distributions, we show that wide tertiaries have isotropic orientations with respect to the inner binaries, and the co-planar orbits can be ruled out. The inferred eccentricity distribution of wide tertiaries is consistent ($<1\sigma$) with being thermal ($f(e)\propto e$), similar to wide binaries at similar separations. The dynamical unfolding scenario is disfavored because it predicts highly eccentric wide tertiaries, which is inconsistent with our findings. For the Kozai-Lidov mechanism to be effective for wide tertiaries at $>10^3$\,AU, the initial separations of the inner binaries need to be $>3$\,AU. Future theoretical investigations are needed to explore the parameter space at these large initial separations and large tertiary separations.

\end{abstract}

\keywords{binaries: general --- binaries: visual --- stars: kinematics and dynamics }

\section{Introduction}

The formation of close binaries has been an open question for decades. Thousands of main-sequence binaries with orbital periods $P<1$\,days (semi-major axes $a\sim0.01$\,AU) have been discovered \citep{Duquennoy1991,Paczynski2006,Prsa2011,Duchene2013,Moe2017,Jayasinghe2020}, but their initial binary separations need to be larger than $a\sim10$\,AU due to the size of the initial hydrostatic stellar core \citep{Larson1969}. Therefore, these close binaries must have experienced a significant orbital migration, shrinking their orbital separations by $\sim3$ orders of magnitudes.

Observations reveal that a large fraction of short-period binaries with $P<$\, a few days have tertiary companions \citep{Tokovinin2006,Pribulla2006,Rappaport2013,Hwang2020c}. This strong tendency to have tertiary companions was originally thought to be the evidence that short-period binaries are formed through the Kozai-Lidov mechanism \citep{Kozai1962,Lidov1962}, where the presence of the tertiary companions may excite high eccentricities of the inner binaries and then the tidal effect can shrink the orbit at the pericenter passage \citep{Harrington1968,Kiseleva1998, Eggleton2001,Fabrycky2007,Naoz2013,Naoz2014}. Furthermore, from the measurements of eclipse timing variation, the distribution of mutual inclination between the orbit of tertiary and that of inner binaries is enhanced at $40^\circ$ \citep{Borkovits2016}, in agreement with the prediction from the Kozai-Lidov mechanism \citep{Fabrycky2007}. This part of the observations supports the hypothesis that close binaries are formed through the Kozai-Lidov mechanism.

However, the Kozai-Lidov scenario is inconsistent with other properties of observed binaries and triples. First, the tendency to have tertiary companions remains true for tertiaries at separations $>10^3$\,AU \citep{Hwang2020c}, where the Kozai-Lidov effect is only effective when the initial separations of inner binaries are large, $>$ a few AU. Furthermore, despite the enhanced mutual inclinations at $40^\circ$, eclipse timing variation also shows that $\sim50$\% of the tertiaries have nearly co-planar (mutual inclinations $<15^\circ$) orbits with respect to the inner binaries \citep{Borkovits2016}, where the Kozai-Lidov effect cannot be excited. These observational results challenge the Kozai-Lidov mechanism as the dominant close binary formation channel. 

Some other formation scenarios have been proposed for close binary formation. For example, a substantial orbital migration may occur during the pre-main-sequence stage through the interaction with the surrounding gas \citep{Moe2018}, and the correlation between close binary and tertiary formation may be established due to the enhanced accretion rates during the formation \citep{Tokovinin2020}. However, for short-period binaries at $P<1$\,days, there is a lack of young binaries with ages $<1$\,Gyr \citep{Hwang2020b}, indicating that their orbital migration takes place on $\sim$Gyr timescales. Therefore, these short-period binaries cannot be formed directly through the Myr-timescale pre-main-sequence interaction. Another formation scenario is the dynamical unfolding of compact triples, where three stars were born in a compact configuration, and the chaotic dynamical evolution may lead to a hierarchical triple if it is not disrupted \citep{Reipurth2012,Elliott2016}.

The all-sky astrometry survey Gaia \citep{Gaia2016} provides a revolutionary dataset for binary stars. With its photometric, spectroscopic, and astrometric capabilities, Gaia covers binaries of various types, including eclipsing binaries, spectroscopic binaries, unresolved astrometric binaries \citep{Belokurov2020, Penoyre2022b, Andrew2022}, and resolved wide binaries \citep{Oh2017, El-Badry2018b, Jimenez-Esteban2019,Tian2020,Hartman2020,El-Badry2021,Hwang2022halo}. These Gaia binaries with separations spanning from $\sim0.01$\,AU to $\sim10^5$\,AU form a golden sample for understanding the formation of binaries and higher-order multiples.

Gaia can further constrain the orientations and eccentricities of wide tertiaries around inner binaries using the so-called $v$-$r$ angles. In a Keplerian orbit, $v$-$r$ angles are the angles between the separation vector ($r$) and the relative velocity vector ($v$), and their projected quantities can be measured by Gaia's high-precision astrometry. $v$-$r$ angles provide a unique approach to infer the eccentricities of wide binaries whose orbital periods are $>10^3$\,year \citep{Tokovinin2016,Tokovinin2020a,Tokovinin2022,Hwang2022ecc}. We applied this method to twin wide binaries and showed that they are highly eccentric \citep{Hwang2022twin}, suggesting that they were formed at smaller separations and their orbits were widened by subsequent interaction with the environments \citep{El-Badry2019}. If one of the component star is an eclipsing binary or has a transiting planet, then $v$-$r$ angles can constrain the orbital alignments between the outer companions and the inner eclipsing or transiting systems \citep{Behmard2022,Christian2022}.

In this paper, we investigate the wide tertiary (at $10^3$-$10^4$\,AU) fraction around inner binaries with different orbital periods. We further study the orientation and eccentricity of the wide tertiaries using their $v$-$r$ angles. The paper is structured as follows. The sample selection is detailed in Sec.~\ref{sec:sample}. We present the wide tertiary fraction in Sec.~\ref{sec:wcf} and the $v$-$r$ angle distributions and the inferred eccentricity in Sec.~\ref{sec:vr}. We discuss the implications for their formation in Sec.~\ref{sec:discussion} and conclude in Sec.~\ref{sec:conclusion}. For triple systems, we use $a_{in}$ ($a_{out}$) and $P_{in}$ ($P_{out}$) to refer to the semi-major axis and the orbital period of the inner binary (outer tertiary companion). 

\section{Sample selection}

\label{sec:sample}

\subsection{Close binary samples}

\label{sec:sample-close}

\begin{figure*}
    \centering
    \includegraphics[width=\linewidth]{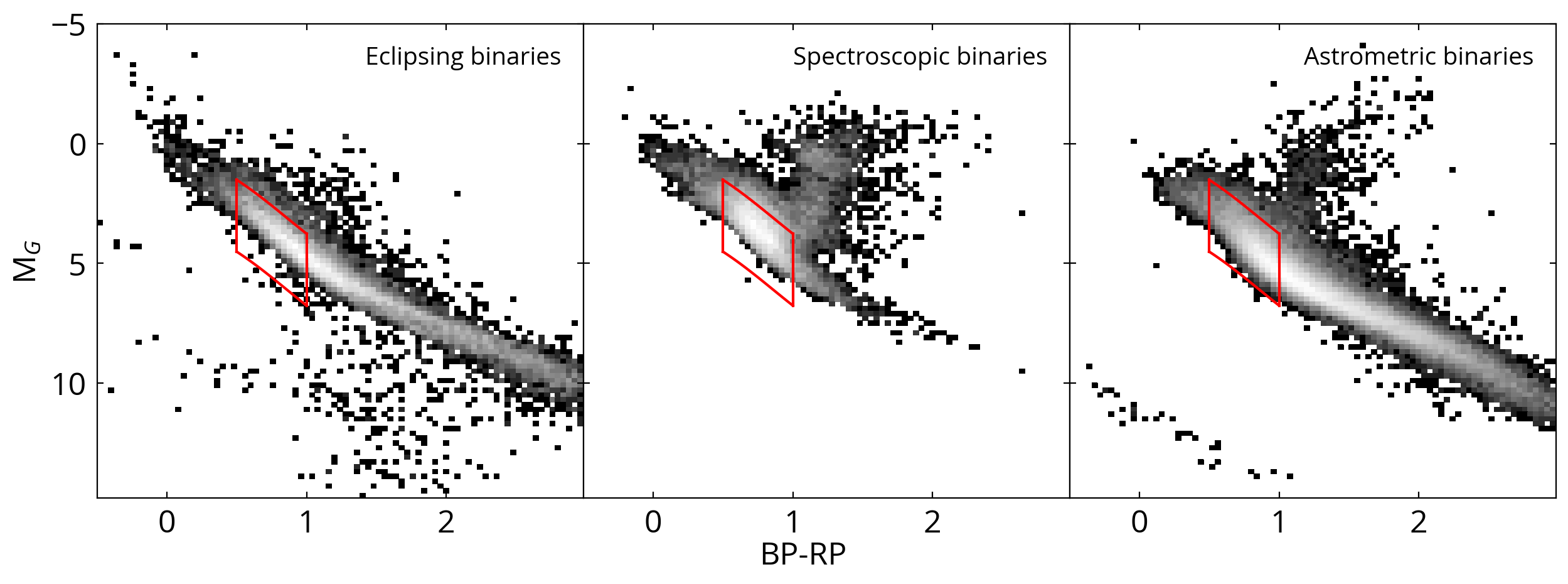}
    \caption{The Hertzsprung-Russell (H-R) diagram for Gaia DR3's eclipsing binaries (left), spectroscopic binaries (middle), and astrometric binaries (right). The color represents the density of the stars on a log scale. Different types of close binaries are sensitive to different regions of the H-R diagram. We select the red-box region as our main sample for the analysis of wide tertiary fractions.  }
    \label{fig:HR}
\end{figure*}

\begin{figure}
    \centering
    \includegraphics[width=\linewidth]{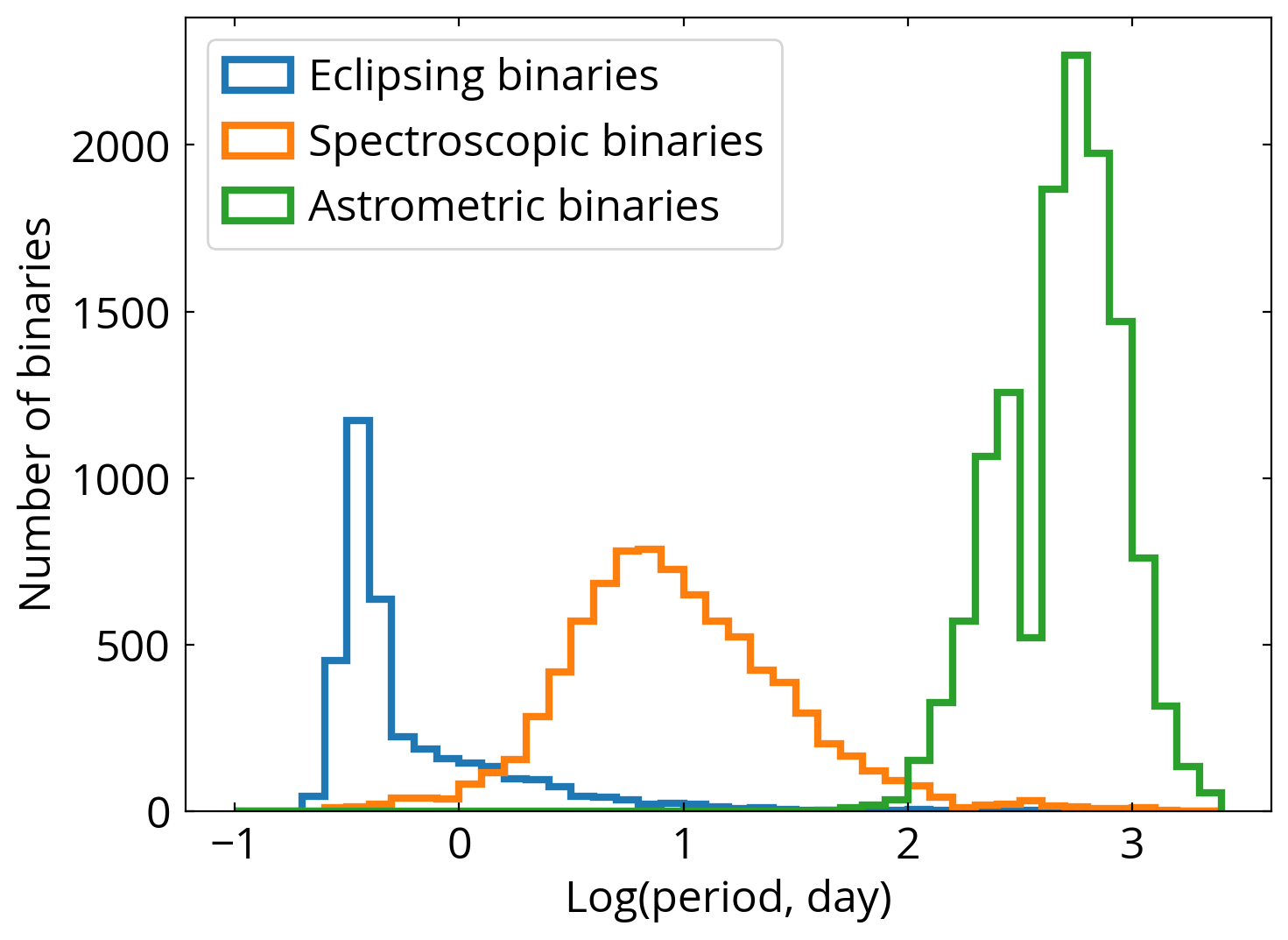}
    \caption{The period distributions of Gaia DR3 close binaries in our sample. The entire close binary sample covers the orbital periods of inner binaries for four order of magnitudes, with eclipsing binaries at $P_{in}\lesssim 1$\,day, spectroscopic binaries at $P_{in}\sim10$\,days, and astrometric binaries at $P_{in}=10^2$-$10^3$\,days. The deficit of astrometric binaries at periods of 1\,year is due to the degeneracy between astrometric solutions and the 1-year parallax motion. }
    \label{fig:P-dist}
\end{figure}

With 34 months of observations, Gaia Data Release 3 (DR3, released on June 13, 2022; \citealt{GaiaDR3Vallenari2022}) provides unprecedented all-sky catalogs for eclipsing, spectroscopic, and astrometric binaries. Eclipsing binaries are identified using Gaia's G-band time series \citep{GaiaDR3Eyer2022}. After period search and modeling of the time series, variable stars are classified by machine-learning-based supervised classification (Remoldini et al. in preparation), resulting in $2.2$ million eclipsing binary candidates in the \texttt{vari\_eclipsing\_binary} table\footnote{\url{https://doi.org/10.17876/gaia/dr.3/82}} (Mowlavi et al. in preparation). These stars are further processed by detailed light curve modeling, and only 4\% of the stars with accepted solutions are included in the \texttt{nss\_two\_body\_orbit} table\footnote{\url{https://doi.org/10.17876/gaia/dr.3/74}} with the column \texttt{nss\_solution\_type=EclipsingBinary} (Siopis et al. in preparation). From the comparison with OGLE-IV \citep{Udalski1992}, the completeness of the \texttt{vari\_eclipsing\_binary} table is 33\%, 45\%, 19\% in the Large Magellanic Cloud, Small Magellanic Cloud, and the bulge region, respectively, and the contamination is $\sim5$\% \citep{GaiaDR3Eyer2022}. In this paper, we use the eclipsing binaries from the \texttt{vari\_eclipsing\_binary} table for its larger sample size than \texttt{nss\_two\_body\_orbit}.

The \texttt{vari\_eclipsing\_binary} table models the G-band time series using up to two Gaussian functions and one sine function \citep[][Mowlavi et al. in preparation]{Mowlavi2017} and 
provides the frequency measurements for the eclipsing binaries, and we compute their periods by $1/$frequency. Ideally, this period is the orbital period of the binary, but there is a well-known problem that period search of eclipsing binaries often finds periods differing from the orbital periods by a factor of two because the primary and secondary eclipses are sometimes less distinguishable (e.g. \citealt{VanderPlas2018, Petrosky2021}). Therefore, we caution that the period may be a factor of two different from the actual orbital period, but this does not affect our main results.

Spectroscopic binaries in Gaia DR3 include single-lined spectroscopic binaries (SB1) and double-lined spectroscopic binaries (SB2). SB1 and SB2 probe similar binary orbital periods from a few days to a few hundred days. SB2 are more sensitive to mass ratios $q\sim1$ where both component stars contribute significantly to the spectra \citep{Kounkel2021}, while SB1 probe smaller mass ratios. In the \texttt{nss\_two\_body\_orbit} table, there are 4630 SB2 and 181327 SB1 (Damerdji et al. in preparation). In this work, our primary focus is the binary period, and thus we include both SB2 and SB1 in the same spectroscopic binary (SB) category, and we expect that SB1 dominates the sample. Following \cite{GaiaDR3Arenou2022}, we require the significance of the primary's radial velocity semi-amplitude larger than 40 to avoid spurious orbital solutions.

Gaia's high-precision astrometry capability enables the identification of unresolved astrometric binaries. These astrometric binaries have orbital periods shorter than or comparable to the 3-year baseline of Gaia DR3, and their light center's motion due to the underlying orbital motion is detectable by Gaia \citep{Penoyre2022a}. Gaia DR3 pre-selects stars where the single-star solutions have large residual errors with \texttt{ruwe}$>1.4$ and processes them with astrometric binary solutions, resulting in 140k sources with orbital period and eccentricity solutions (\texttt{nss\_solution\_type}=\texttt{Orbital}) in the \texttt{nss\_two\_body\_orbit} table \citep{GaiaDR3Halbwachs2022}. This table provides updated parallax and proper motion measurements resulting from the astrometric binary solutions, in contrast to the single-star solution in the main table (\texttt{gaiadr3.gaia\_source}). We use the updated parallaxes and proper motions for astrometric binaries in the analysis.

Some stars have multiple entries in the \texttt{nss\_two\_body\_orbit} table because their close companions are detected by multiple methods (e.g. \texttt{nss\_solution\_type=SB1} and \texttt{Orbital}). It can be the same companion detected by multiple methods, or can be different companions where the periods and eccentricities are inconsistent from different methods. Since these stars with multiple entries are rare and only constitute 3\% of the \texttt{nss\_two\_body\_orbit} table, we do not include them in the analysis. For the same reason, we exclude sources in the \texttt{vari\_eclipsing\_binary} table that have entries in the \texttt{nss\_two\_body\_orbit} table with \texttt{nss\_solution\_type} other than \texttt{EclipsingBinary}. 

In the analysis of wide tertiary fractions in Sec~\ref{sec:wcf}, we focus on the binaries with parallaxes $>2$\,mas (and thus distances $<500$\,pc) and parallax over error $>10$. These parallax criteria ensure that wide tertiaries at $>10^3$\,AU are $>2$\,arcsec from the other source, and thus their detection completeness is high and the wide tertiaries do not strongly affect the BP/RP flux measurements of the close binaries. For all close binaries in the analysis, we require their \texttt{phot\_g\_mean\_flux\_over\_error}$>10$. In the analysis where the BP and RP of close binaries are used, we require \texttt{phot\_bp\_mean\_flux\_over\_error}$>10$,  \texttt{phot\_rp\_mean\_flux\_over\_error}$>10$, and \texttt{phot\_bp\_rp\_excess\_factor}$<$1.8, where the last criterion is to ensure that BP and RP are not strongly affected by crowdedness \citep{Evans2018}.

Fig.~\ref{fig:HR} shows the Hertzsprung-Russell (H-R) diagrams for eclipsing binaries (left), spectroscopic binaries (middle), and astrometric binaries (right) selected by the criteria detailed above. Different binary types are sensitive to different parts of the H-R diagram. To have similar primary masses among three close binary samples, we use the red boxes in Fig.~\ref{fig:HR} to select binaries in the same main-sequence region of the H-R diagram. Specifically, the red box is defined by (1) BP-RP colors between 0.5 and 1; and (2) the difference of absolute G-band magnitudes $<1.5$\,mag from the Pleiades' main sequence fit \citep{Hamer2019}. The sample within this region is dominated by main-sequence stars, and the $<1.5$\,magnitude difference ensures that unresolved binaries are selected (unresolved binaries are $<0.75$\,mag brighter than single stars). This main-sequence selection roughly corresponds to a mass range between 1.4 and 0.8\Msun \citep{Paxton2011,Dotter2016,Choi2016}. These selections result in 3640 eclipsing binaries, 8433 spectroscopic binaries, and 12835 astrometric binaries in our analysis. 

Using the same selection criteria and the main-sequence cut, we query a field star sample of 1.5 million stars within 500\,pc. 93\% of them have \texttt{non\_single\_star=0} in the  \texttt{gaiadr3.gaia\_source} table, meaning that they are not considered as non-single stars (including eclipsing, spectroscopic, and astrometric binaries) in the \texttt{nss\_two\_body\_orbit} table. Correcting for the completeness of close binary detectability, \cite{Moe2017} report a close binary fraction at $P_{in}<10^{3.7}$\,days and mass ratios $q>0.1$ of $15\pm3$\% for the solar-type stars. In this paper, most of the close binaries in Gaia DR3 have orbital periods $P_{in}<10^3$\,days (Fig.~\ref{fig:P-dist}), and therefore this field star sample serves as a control sample where most ($>85$\%) of the stars are not close binaries.

Fig.~\ref{fig:P-dist} shows the period distributions of the resulting main-sequence close binaries. Eclipsing binaries (blue) are strongly clustered at periods $<1$\,day because binaries with smaller orbits have a larger probability of being eclipsed. The periods of spectroscopic binaries span from $\sim$1 to $\sim10^2$\,days and peak around 10\,days. The astrometric binaries have orbital periods ranging from $10^2$ to $10^3$\,days, where Gaia DR3's baseline sets the upper limit. The deficit of astrometric binaries at periods of 1\,year is due to that binaries' orbital motion with a 1-year period is difficult to be decoupled from the parallax motion, which also has a period of 1\,year. The median periods are 0.41, 8.42, and 533 days for eclipsing, spectroscopic, and astrometric binaries, respectively. With Gaia's multiple close binary samples, we can investigate their properties across four orders of magnitude in orbital periods, corresponding to inner semi-major axes from $a_{in}\sim0.01$ to $\sim1$\,AU.

\subsection{Wide tertiary and wide binary samples}

\label{sec:sample-wide}

In this paper, we use the wide binary catalog from Gaia early DR3 \citep[EDR3,][]{El-Badry2021}, where wide binaries are searched out to 1-kpc distance from the Sun with binary separations up to 1\,pc. Note that the photometric and astrometric content of Gaia DR3 is nearly unchanged from Gaia EDR3 \citep{GaiaDR3Vallenari2022}. The resolved wide binaries (or wide tertiaries around unresolved close binaries) are identified by their small proper motion differences that are consistent with the Keplerian motion. 

The wide binary catalog from \cite{El-Badry2021} uses the single-star astrometric solutions from Gaia EDR3. However, for astrometric binaries, their orbital motions induce significant astrometric noise in the single-star solutions, which may affect their wide tertiary search. Therefore, using the public codes\footnote{\url{https://doi.org/10.5281/zenodo.4435257}} from \cite{El-Badry2021}, we rerun the wide tertiary search for astrometric binaries with updated parallaxes and proper motions from their non-single-star solutions in the \texttt{nss\_two\_body\_orbit} table. All selection criteria for wide tertiaries are identical to the EDR3 wide binary catalog.

We find that for astrometric binaries at $P_{in}<10^3$\,days within 500\,pc, the number of wide tertiaries from the original EDR3 catalog is $>80$\% than the number from our rerun. This fraction drops significantly to 45\% at $P_{in}>10^3$\,days, suggesting that the orbital motions of these longer-period astrometric binaries strongly affect their single-star solutions in EDR3 and reduce the number of the resulting wide tertiaries.

For astrometric binaries in the following analysis, we use the wide tertiaries from our rerun to account for their non-single-star solutions. \cite{El-Badry2021} also estimate the probability of being a chance-alignment pair (\texttt{R\_chance\_align}) for each wide binary. While we are not able to re-compute \texttt{R\_chance\_align} for the new wide tertiaries identified from our rerun, we find that most of the wide tertiaries with separations $<10^4$\,AU are unlikely to be chance-alignment pairs. For example, applying \texttt{R\_chance\_align}$<0.1$ to the original EDR3 wide binary catalog only affects the wide ($10^3$-$10^4$\,AU) tertiary fraction less than 0.1\%. Therefore, in the main analysis, we do not apply the \texttt{R\_chance\_align} criterion.

\begin{figure}
    \centering
    \includegraphics[width=\linewidth]{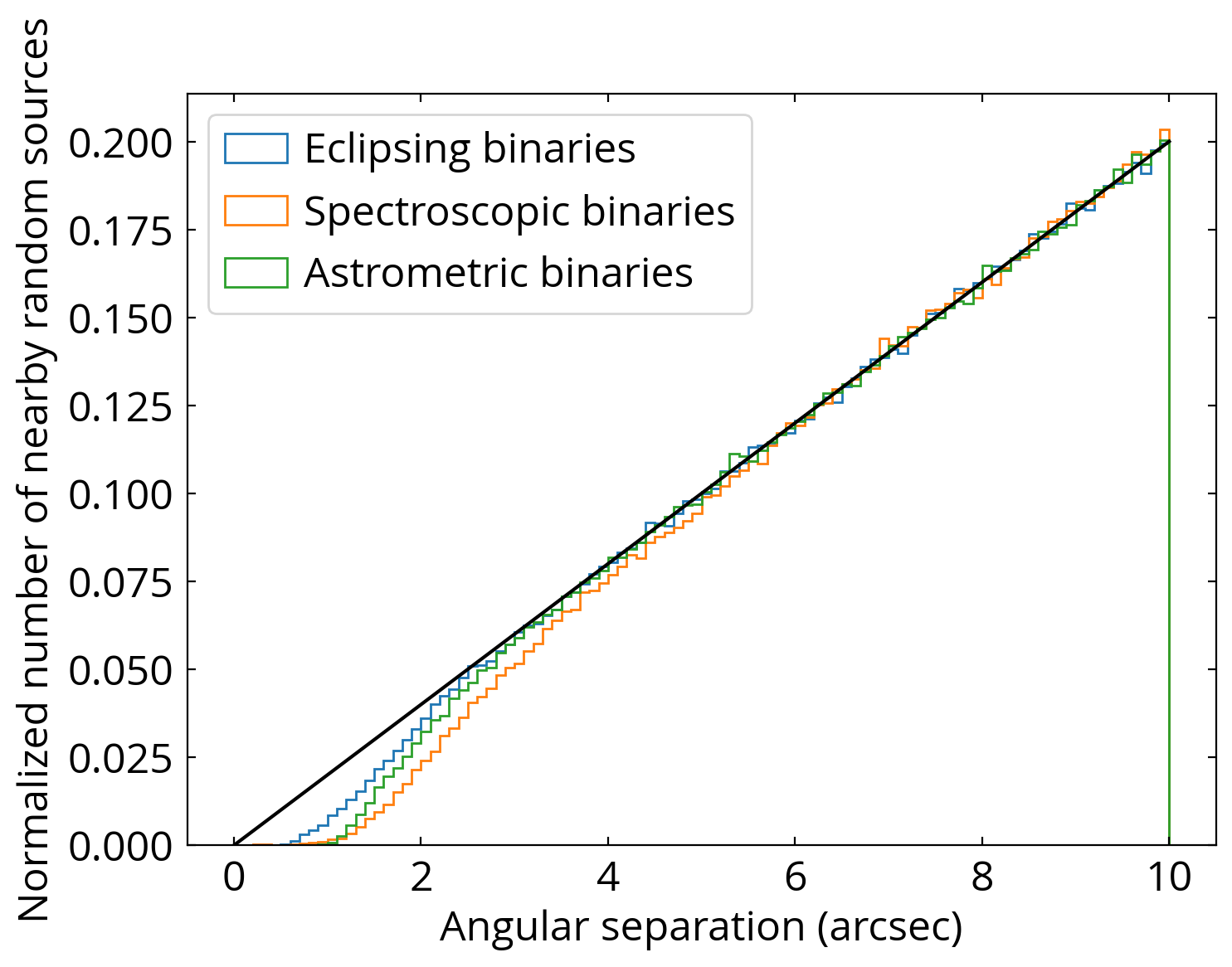}
    \includegraphics[width=\linewidth]{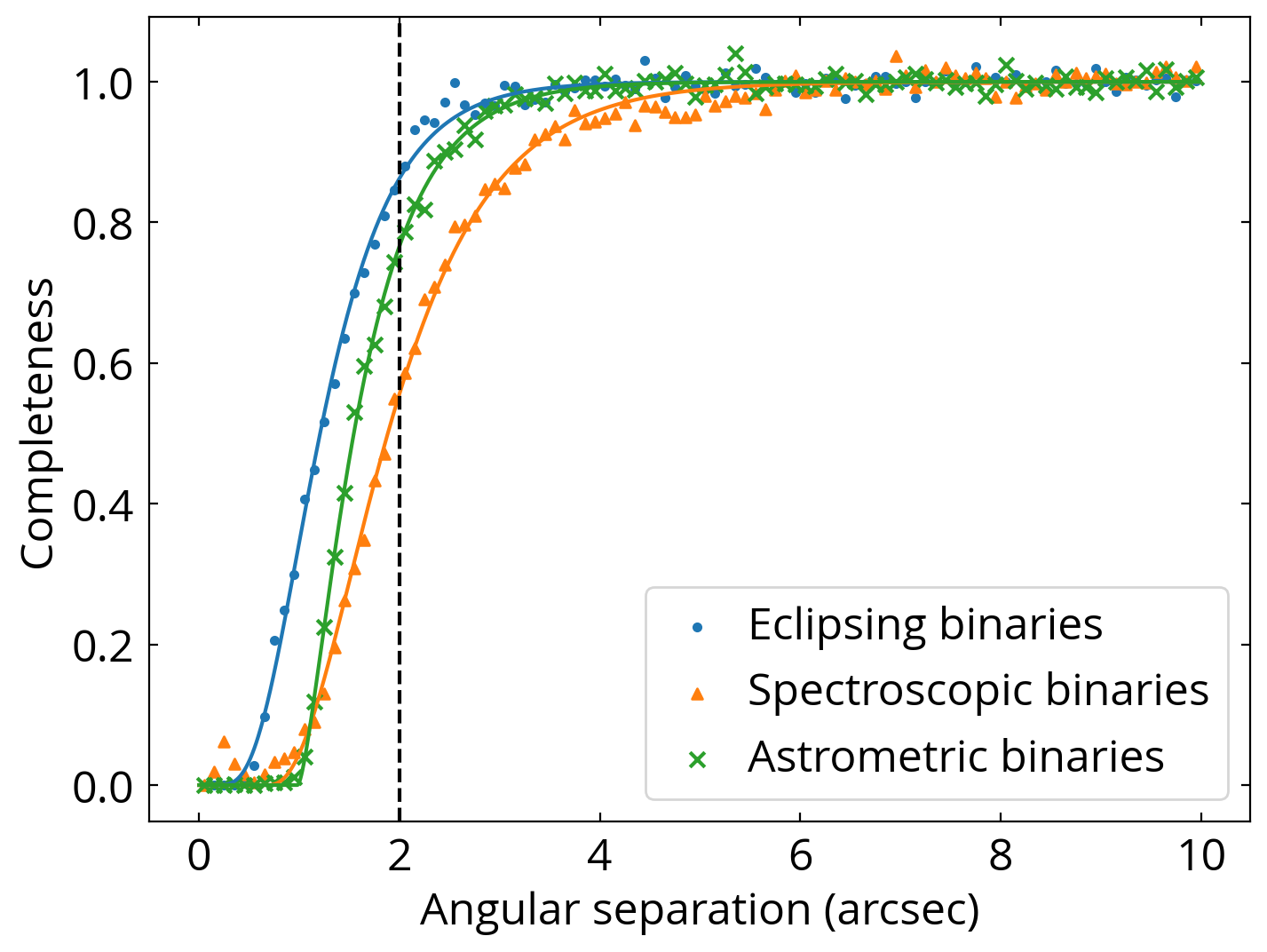}
    \caption{The completeness of nearby sources around Gaia DR3 close binaries. Top: angular separation distributions of nearby sources around different types of close binaries. The distributions are normalized at 5-10\,arcsec, where they agree with the expected distribution for chance-alignment pairs (black line). Bottom: the completeness of nearby sources around different types of Gaia close binaries. All three categories have completeness $>0.56$ at $>2$\,arcsec (dashed vertical line). }
    \label{fig:completeness}
\end{figure}

\subsection{Completeness of nearby sources around Gaia close binaries}

\label{sec:completeness}

The presence of nearby stars may affect the detectability of close binaries. The wide binary catalog, which does not involve any close binary identifications, has better angular completeness; therefore, its completeness is not the limiting factor \citep{El-Badry2021}. Since one of the goals in this paper is to quantify the wide tertiary fraction around close binaries, it is critical to quantify the completeness of nearby sources around Gaia close binaries.

To test the completeness, we collect all pairs within 10\,arcsec around eclipsing binaries, spectroscopic binaries, and astrometric binaries. The Gaia query used to search for nearby sources around close binaries is provided in Appendix~\ref{sec:appendix-query}, where we use the \texttt{non\_single\_star} column in the main table (\texttt{gaiadr3.gaia\_source}) to identify close binaries. The eclipsing binaries in this table are only a subset of the \texttt{vari\_eclipsing\_binary} table, but their completeness properties are expected to be similar. For completeness correction here, we do not impose parallax nor color cuts on the pair sample. By comparing with the Gaia EDR3 wide binary catalog \citep{El-Badry2021}, we find that at most $10\%$ of these pairs are physical wide binaries with distances $<1$\,kpc, and this fraction is highest at $<2$\,arcsec and for spectroscopic binaries. We exclude these known wide binaries from the pair sample. The presence of wide binaries with distances $>1$\,kpc distances is possible and may make us underestimate the completeness, but due to the steep binary separation distribution \citep{El-Badry2018b} and to that the faint binary companions may fall out of Gaia's detection limit, such contribution is at most a few percent at $<2$\,arcsec in the pair sample, and even lower at $>2$\,arcsec for our main analysis.

Fig.~\ref{fig:completeness} top panel shows the separation distributions of nearby pairs for three categories of close binaries. These separation distributions are normalized at 5-10\,arcsec. The observed distributions at $\gtrsim3$\,arcsec agree with the expected distribution for chance-alignment pairs $N(s)ds\propto s$ where $s$ is angular separation (black line), supporting the fact that the sample is dominated by chance-alignment pairs. All three categories of close binaries show a deficit of nearby sources below $\sim3$\,arcsec, larger than Gaia's pair completeness of down to $\sim0.5$\,arcsec when no close binary identifications are involved \citep{Gaia2021Fabricius}. Therefore, Gaia's close binary detectability is affected by nearby sources at $\lesssim3$\,arcsec.

To correct for completeness, we derive the completeness by the ratio of the observed angular separation distribution to the expected distribution (black line in the top panel). The bottom panel of Fig.~\ref{fig:completeness} shows the resulting completeness. The completeness around eclipsing binaries (blue) and astrometric binaries (green) is $>0.76$ at $2$\,arcsec and $>0.96$\,at $3$\,arcsec. Spectroscopic binaries have a lower completeness, with 0.56 at 2\,arcsec and $>0.86$ at $>3$\,arcsec.

We fit a functional form of $C(s)=(1-exp(-A(s-s_0)))^B$ when the angular separation $s>s_0$ and $C(s)=0$ when $s<s_0$. The best-fit parameters are $s_{0,EB}=0.12$, $A_{EB}=1.88$, $B_{EB}=5.00$, $s_{0,SB}=0.66$, $A_{SB}=1.29$, $B_{SB}=2.97$, $s_{0,AB}=0.98$, $A_{AB}=1.81$, $B_{AB}=1.54$ for three categories of close binaries (EB, SB, and AB are eclipsing, spectroscopic, and astrometric binaries, respectively). The best fits are shown as solid lines in the bottom panel of Fig.~\ref{fig:completeness}.

When computing the wide tertiary fraction, we use 2\,arcsec as the limiting angular separations. For stars at distances $<500$\,pc, 2\,arcsec corresponds to the tertiary separations at $>1000$\,AU. To account for the completeness in the calculation of the wide tertiary fraction, we assign weights $=1/C(s)$ for each pair based on its angular separation $s$. Typically the completeness correction affects our measurements by an amount smaller than their uncertainties, thus playing a minor role in our results.

\begin{figure}
    \includegraphics[width=\linewidth]{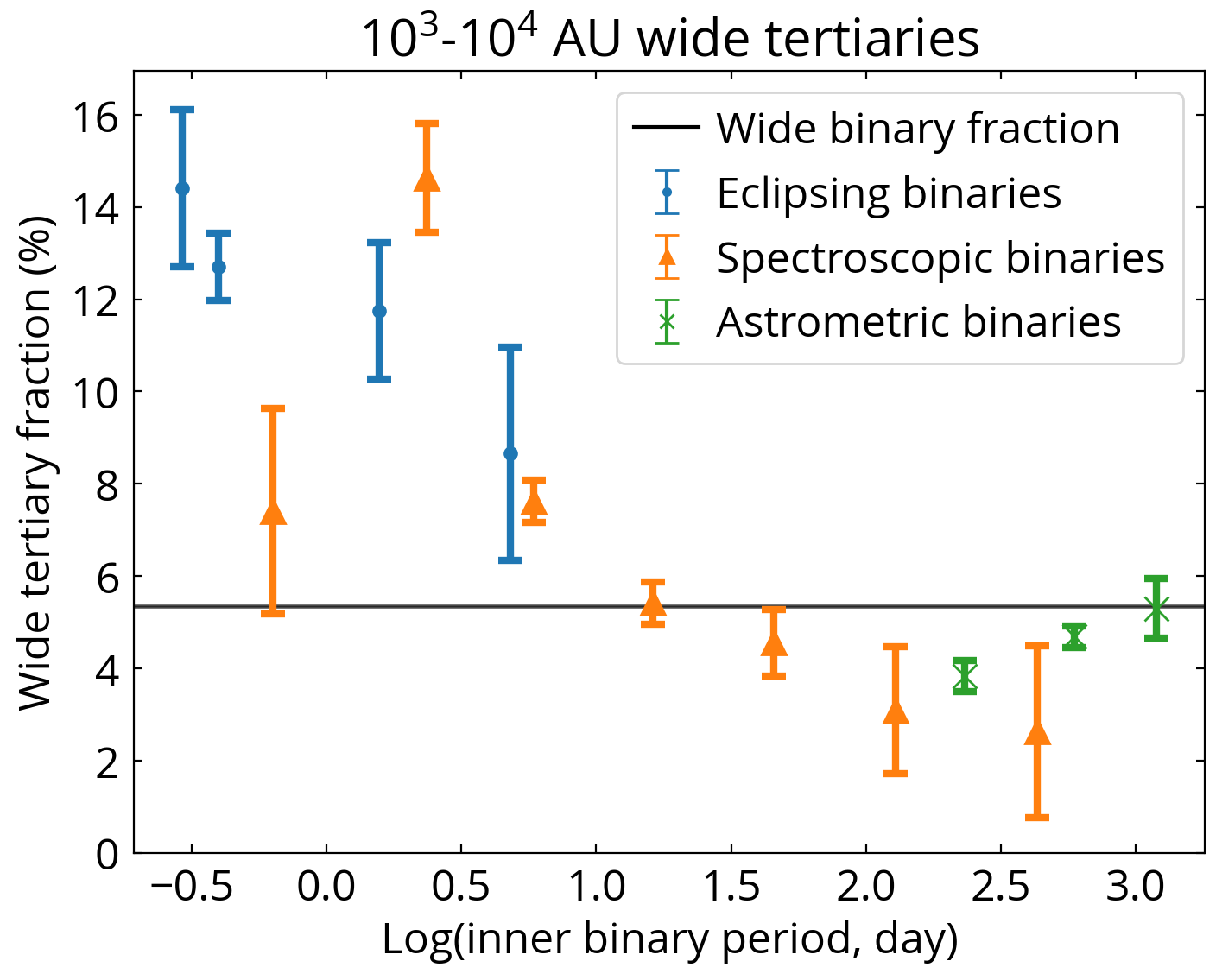}
    \caption{The wide tertiary fractions as a function of inner binaries' orbital periods. The black horizontal lines represent the wide binary fraction of the field stars, with the breadth indicating the small uncertainty. The wide tertiary fractions strongly increase with decreasing inner binaries' orbital periods. }
    \label{fig:WCF-P}
\end{figure}

\begin{figure*}
    \includegraphics[width=0.5\linewidth]{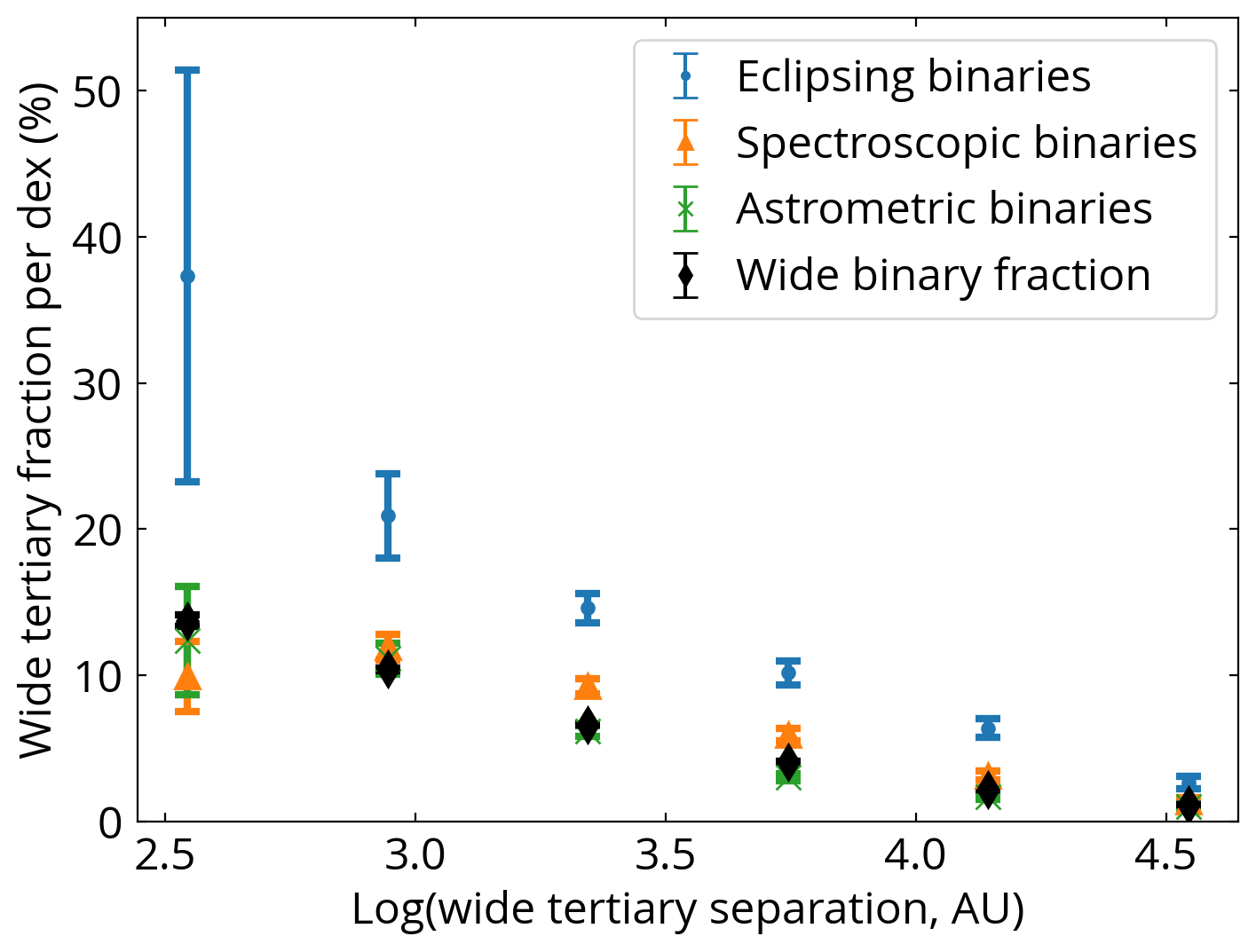}
    \includegraphics[width=0.5\linewidth]{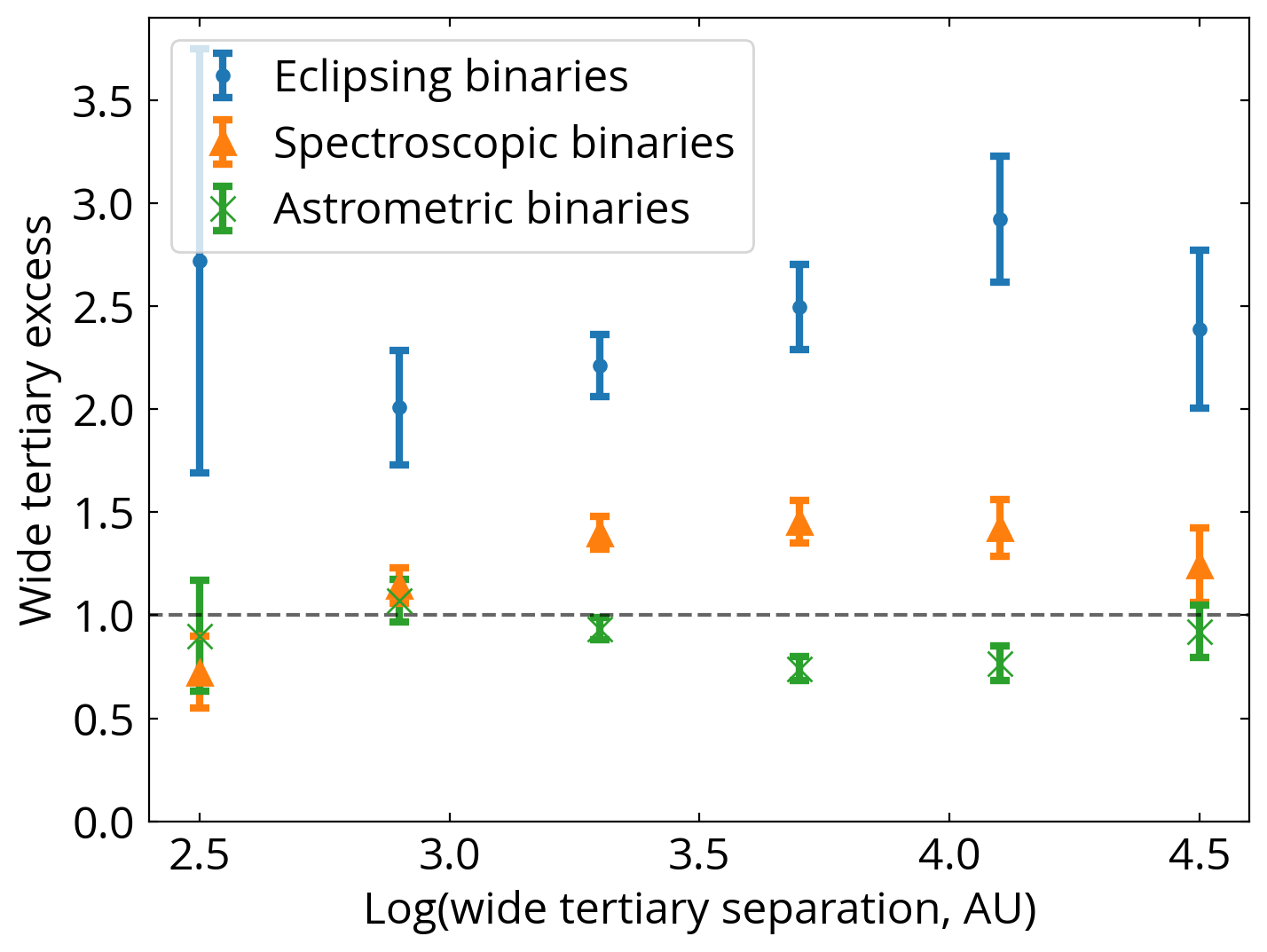}
    \caption{Left: the wide tertiary fraction as a function of tertiary separations. The black diamonds show the wide binary fraction versus binary separations. Similar to wide binaries, the wide tertiary fraction increases toward smaller separations. Right: the wide tertiary excess as a function of tertiary separations, where the wide tertiary excess is the ratio of tertiary fractions to the wide binary fraction at the same separation. Both eclipsing and spectroscopic binaries show a tentative increasing wide tertiary excess with respect to tertiary separations.
 }
    \label{fig:WCF-sep}
\end{figure*}

\section{Wide tertiary fraction}

\label{sec:wcf}

Fig.~\ref{fig:WCF-P} shows the wide tertiary fraction as a function of inner binaries' orbital periods. A wide tertiary fraction is the fraction of close binaries that have wide tertiaries in a certain range of projected separations. Within 500-pc, our wide tertiaries are complete down to absolute G-band magnitudes of $\sim12$\,mag, covering most of the stars except for white dwarfs and the faint end of M dwarfs. Given the high completeness of the tertiary companions, we do not apply further selections to them (i.e. no color and flux-over-error selections). We consider wide tertiaries at $10^3$-$10^4$\,AU in Fig.~\ref{fig:WCF-P} and other separations in Appendix~\ref{sec:appendix}. The black horizontal line is the wide binary fraction of the field stars, and its breadth indicates the 1-$\sigma$ uncertainty. The bin size of the inner binary periods is from $10^{-1}$ to $10^{3.5}$\,days with a step of 0.5\,dex. Symbols' x-axis values are the median inner binary periods in each bin. The error bars are the 1-$\sigma$ Poisson-counting uncertainties in wide tertiary fractions, and we discard measurements with uncertainties $>3$ \%.

Fig.~\ref{fig:WCF-P} shows that the wide tertiary fraction increases with decreasing inner binaries' orbital periods, with the wide tertiary fraction higher than the wide binary fraction at $P_{in}<10$\,days. Specifically, the wide tertiary fraction is $4.53\pm0.19$\% for entire astrometric binaries (median $P_{in}=533$\,days), $7.35\pm0.30$\% for spectroscopic binaries (median $P_{in}=8.42$\,days), and $12.48\pm0.59$\% for eclipsing binaries (median $P_{in}=0.41$\,days). The wide tertiary fraction of eclipsing binaries is $2.33 \pm 0.11$ times higher than the wide binary fraction ($5.35\pm0.02$\%), in agreement with the finding from \cite{Hwang2020c}. \cite{Tokovinin2006} report that 96\% of close binaries at $P_{in}<3$\,days have tertiary companions (at any separations) and this fraction drops to 34\% at $P_{in}>12$\,days, similar to the overall trend in Fig.~\ref{fig:WCF-P} except that we now cover the inner periods for four orders of magnitudes.

All the close binary samples and the field star sample are selected to have similar primary masses using their locations in the H-R diagram (Fig.~\ref{fig:HR}). One may argue that the total masses of the close binaries are more massive than the single stars at the same colors, and thus a better comparison sample might be field stars with masses similar to the total masses of the close binaries. To test this argument, we select bluer main-sequence field stars with BP-RP between 0 and 0.5\,mag,  corresponding to main-sequence masses between 2.1 and 1.4\Msun. Compared to the field stars with BP-RP between 0.5 and 1\,mag, the blue field-star sample has 1.5 times higher masses but only 10\% higher wide tertiary fraction. Therefore, the $2.33\pm0.11$ higher wide tertiary fraction of eclipsing binaries cannot be explained by the larger total masses of close binaries than field stars.

Interestingly, the wide tertiary fractions of spectroscopic and astrometric binaries at $P_{in}\sim10^{2.5}$\,days are lower than the wide binary fraction. Then at $P_{in}=10^3$\,days, the wide tertiary fraction of astrometric binaries becomes consistent with the wide binary fraction. The lower wide tertiary fraction at $P_{in}\sim10^{2.5}$\,days is at 4.5-$\sigma$ significance for astrometric binaries, and the result is nearly unchanged if we use the original Gaia EDR3 wide binary catalog. We discuss some physical possibilities in Sec.~\ref{sec:discussion}, bearing in mind that this lower tertiary fraction can also be due to some Gaia systematics associated with the inner binary's astrometric motion that is not yet well characterized. 

The left panel in Fig.~\ref{fig:WCF-sep} considers the wide tertiary fraction as a function of the projected separations of wide tertiaries. The separation bin starts from $10^{2.3}$\,AU with a step of 0.4\,dex. The vertical axis shows the wide tertiary (binary) fraction per wide tertiary (binary) separation dex. We impose an additional minimum parallax criterion in each bin to ensure that the wide tertiaries in every bin always have angular separations $>2$\,arcsec. The left panel shows that the wide tertiary fraction increases toward smaller separations among all categories, similar to the overall behavior of the wide binary fraction (black diamonds). The right panel presents the wide tertiary excess, the ratio of the wide tertiary fraction to the wide binary fraction at the same separation. Intriguingly, the tertiary excess of eclipsing binaries tentatively peaks around $10^4$\,AU, a particularly interesting separation above which the Galactic tide becomes important \citep{Jiang2010a,Hamilton2022}. The wide tertiary excess of spectroscopic binaries slightly increases from $10^{2.5}$ to $10^{3.5}$\,AU, and that of astrometric binaries is overall flat with slightly lower values around $10^4$\,AU. For wide tertiaries at $10^3$-$10^4$\,AU, the wide tertiary excess is $2.33\pm0.11$, $1.37\pm0.06$, and $0.92\pm0.04$ for eclipsing, spectroscopic, and astrometric binaries, respectively.

It is a mystery how close binaries at $P_{in}<10$\,days ($a_{in}\lesssim10^{-1}$\,AU) have a higher chance of having tertiaries out to $a_{out}=10^4$\,AU, more than five orders of magnitude difference in separations. It seems that the wide tertiaries somehow are aware of the existence of a close binary companion, despite the wide tertiary separations. The wide tertiary excess is $(N_{close+wide}/N_{close})/(N_{field+wide}/N_{field})$, where $N_{close}$, $N_{close+wide}$, $N_{field}$, $N_{field+wide}$ are the numbers of close binaries, close binaries with wide (tertiary) companions, field stars, and field stars with wide (binary) companions, respectively. The wide tertiary excess is equivalent to $(N_{close+wide}/N_{field+wide})/(N_{close}/N_{field})$, the ratio of the close binary fraction in the wide physical pairs to the close binary fraction in the field stars, or the close binary excess in wide pairs. Hence, an alternative interpretation is that close binaries with $P_{in}<10$\,days are more likely to exist in the presence of wide companions. To further understand the nature of the wide tertiaries, we use the $v$-$r$ angles to investigate their orientations with respect to the inner binaries and eccentricities in the next section.

\section{Orientations and Eccentricities of wide tertiaries}
\label{sec:vr}

The $v$-$r$ angles, the angle between the separation vector ($r$) and the relative velocity vector ($v$), provide a unique constraint on the orientation and the eccentricity of the resolved wide tertiaries. If the inner binary has a nearly edge-on orientation (e.g. eclipsing binaries), then the wide tertiary with an aligned co-planar orbit would have $v$-$r$ angles close to 0$^\circ$ or 180$^\circ$ \citep{Behmard2022}. If the orientation of the wide tertiaries is isotropic with respect to the Sun, which can be due to either the isotropic orientation of wide tertiaries with respect to the inner binaries or to the relatively isotropic selection of the inner binaries' orientation (e.g. astrometric binaries), then we can infer the eccentricity distribution from the observed $v$-$r$ angle distribution \citep{Hwang2022ecc}. For example, circular orbits would have a $v$-$r$ angle distribution peaking at $90^\circ$, more eccentric orbits would have peaks moving towards 0$^\circ$ and 180$^\circ$, and the thermal eccentricity distribution ($f(e)\propto e$) corresponds to a uniform $v$-$r$ angle distribution.

In Gaia data, the projected $v$-$r$ angles, the angle between projected $v$ and projected $r$ on the sky, can be measured from the coordinate difference (which is parallel to projected $r$) and the proper motion difference (which is parallel to projected $v$). All projected $v$-$r$ angle measurements for Gaia EDR3 wide binaries are made public from \cite{Hwang2022ecc}, and here we update the $v$-$r$ angle measurements for the wide tertiaries of astrometric binaries from our wide companion search (Sec.~\ref{sec:sample-wide}) using their updated proper motions from the astrometric binary solution. 

In the $v$-$r$ angle analysis, we have different selection criteria than those in Sec.~\ref{sec:wcf}. First, to measure reliable $v$-$r$ angles, we require the significance of the non-zero proper motion differences of the wide pairs be $>3$-$\sigma$. Second, we require parallaxes $>5$\,mas (thus distances $<200$\,pc) and binary separations $<10^4$\,AU to have less biased $v$-$r$ angle distributions. Third, to avoid Gaia's systematics, only pairs with angular separations $>1.5$\,arcsec are included in the analysis. Since the nearby-source completeness (Sec.~\ref{sec:completeness}) plays a minor role here, we do not require the 2-arcsec separation criterion used in Sec.~\ref{sec:wcf}. To improve the sample sizes, we do not impose the main-sequence cut described in Sec.~\ref{sec:sample-close}, so the criteria related to BP and RP bands are not imposed here. We are cautious that without the main-sequence selection, the three close binary categories may have different primary masses. These selections result in 246, 465, and 838 eclipsing, spectroscopic, and astrometric binaries with wide tertiaries, respectively. The details of the systematics tests and selection criteria are discussed in \cite{Hwang2022ecc}.

\begin{figure}
    \centering
    \includegraphics[width=\linewidth]{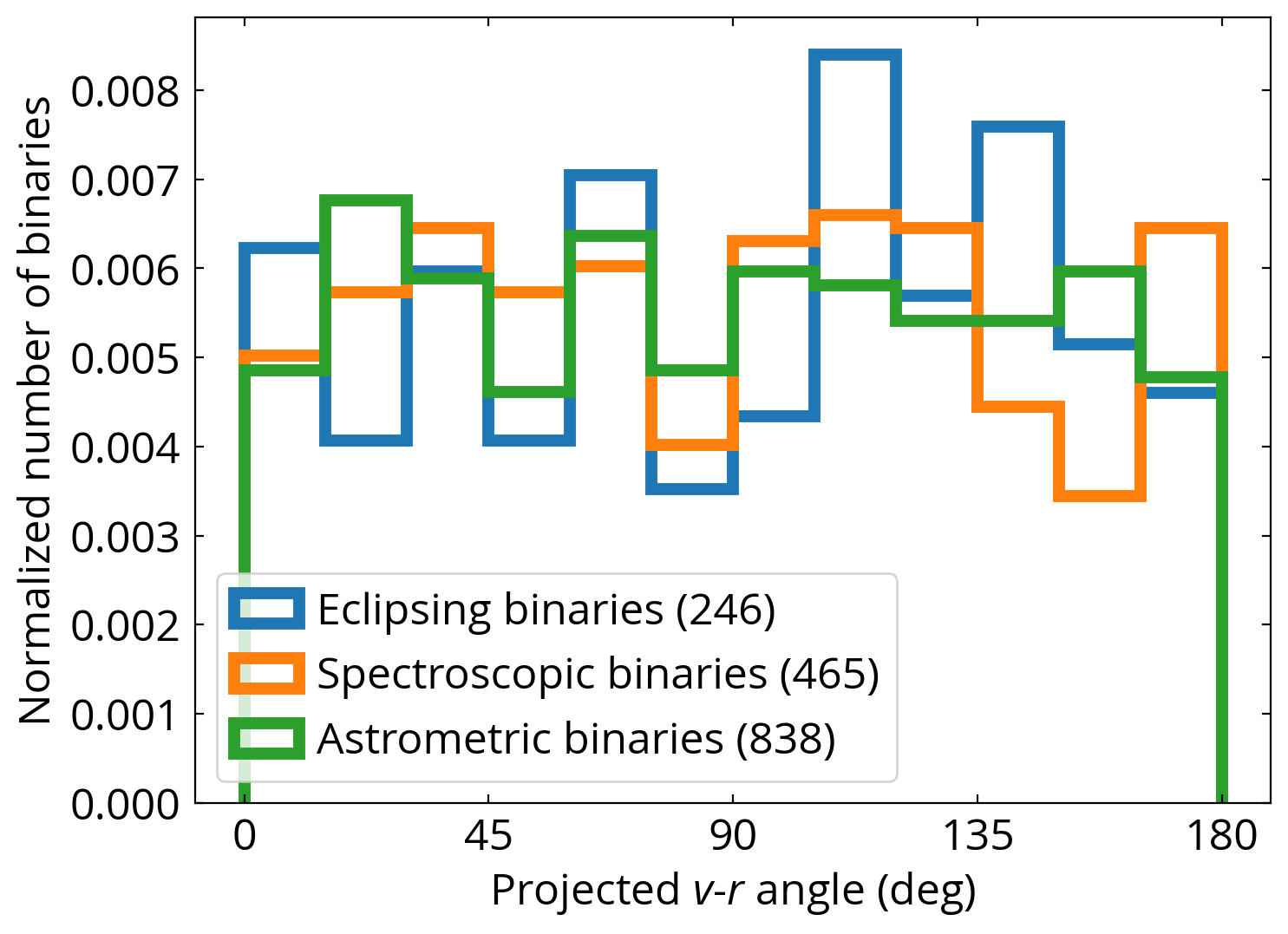}
    \caption{The $v$-$r$ angle distributions of wide tertiaries around different types of close binaries. The parentheses in the legend show the number of sources in each category. Overall their $v$-$r$ angle distributions are consistent with being a uniform distribution. }
    \label{fig:vr}
\end{figure}

Fig.~\ref{fig:vr} shows the $v$-$r$ angle distribution of the wide tertiaries around Gaia close binaries. Using the Kolmogorov-Smirnov (K-S) test, we find that all three $v$-$r$ angle distributions are consistent with each other and are also consistent with the uniform distribution, with a minimum $p$-value of 0.15 among all the tests.  Both eclipsing binaries and spectroscopic binaries are more sensitive to edge-on binaries, and astrometric binaries have relatively flat dependence on the binary orientation with some deficit of edge-on orientations \citep{GaiaDR3Arenou2022}. Therefore, if the wide tertiaries have a strong preference for orientation relative to the inner binaries, then such a signal in the $v$-$r$ angle distribution would be strongest in eclipsing binaries because of the strong selection function of the inner binaries' edge-on orientation. On the other hand, the preferred orientation of the wide tertiaries, if any, would be reduced in the $v$-$r$ angle distribution for astrometric binaries because of the more isotropic selection on inner binaries' orientation. Therefore, the fact that the $v$-$r$ angle distributions are consistent among three close binary categories suggests that the wide tertiaries have isotropic orientations with respect to the inner binaries.

\begin{figure}
    \centering
    \includegraphics[width=\linewidth]{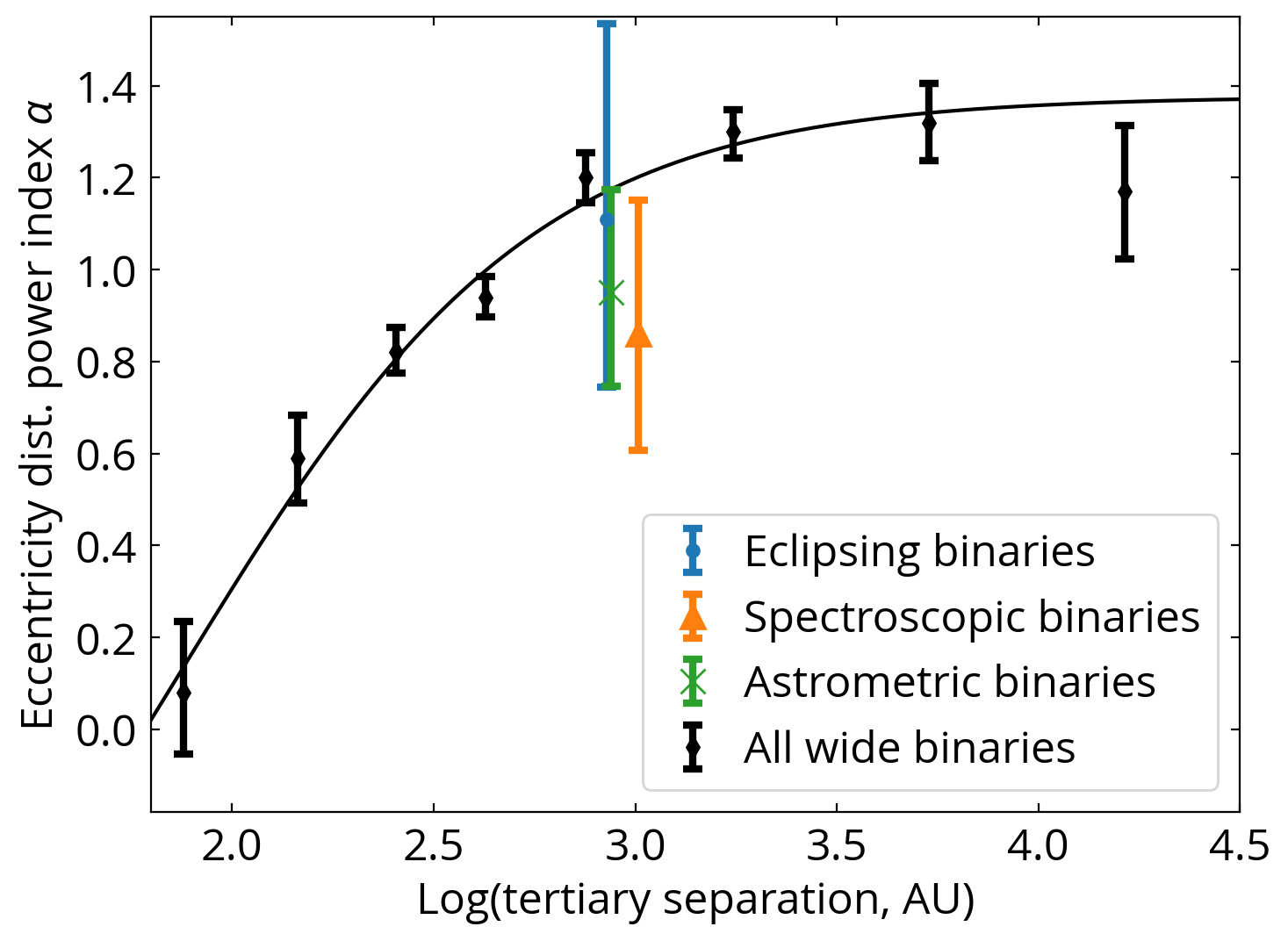}
    \caption{The inferred power-law indices $\alpha$ for the eccentricity distributions ($f(e)\propto e^\alpha$) of wide tertiaries around Gaia close binaries. For comparison, the black points and the solid black line are the results for typical wide binaries where most do not have inner close binaries \citep{Hwang2022ecc}. }
    \label{fig:alpha}
\end{figure}

In Fig.~\ref{fig:alpha}, we translate the $v$-$r$ angle distributions to eccentricity distributions. Specifically, we model the eccentricity distribution using a power law, i.e. $f(e)\propto e^\alpha$, where $e$ is eccentricity. The only free parameter is $\alpha$, and $\alpha=0$ corresponds to the uniform eccentricity distribution and $\alpha=1$ is the thermal eccentricity distribution. In \cite{Hwang2022ecc}, we developed a Bayesian inference framework to derive the best-fit $\alpha$ from the projected $v$-$r$ angle distribution, assuming that the orientation is isotropic. We then applied this method to the Gaia EDR3 wide binaries \citep{El-Badry2021}, with an additional criterion of parallax$>5$\,mas to reduce selection effects. We found that wide binaries have an eccentricity distribution close to uniform at $\sim100$\,AU. Then the eccentricity distribution becomes superthermal ($\alpha>1$) at separations $>10^3$\,AU \citep{Hwang2022ecc}. These results are shown as black markers and solid black line in Fig.~\ref{fig:alpha} for comparison. We note that there is a twin wide binary population that has an excess of highly eccentric ($e>0.95$) orbits \citep{Hwang2022twin}, but at separations $>100$\,AU, they only constitute $<10$\% of the entire wide binary sample \citep{El-Badry2019}. Therefore the black markers in Fig.~\ref{fig:alpha} are dominated by non-twin wide binaries, and excluding twin wide binaries has negligible effects on the results. 

We apply the Bayesian method to wide tertiaries around Gaia close binaries to derive the best-fit power-law index $\alpha$ for the eccentricity distribution. Their measurements are shown in Fig.~\ref{fig:alpha}, where their values on the horizontal axis are the median projected separations of the samples and the vertical error bars are the 68\% credible interval. The inferred $\alpha$ are $1.11_{-0.37}^{+0.42}$, $0.86_{-0.25}^{+0.29}$, $0.95_{-0.20}^{+0.22}$ for the wide tertiaries of the eclipsing, spectroscopic, and astrometric binaries, respectively. These inferred $\alpha$ are consistent within $\sim1$-$\sigma$ among different close binary samples and the typical wide binary sample.

Using the $v$-$r$ angle distribution, we find that the wide tertiaries have an isotropic orientation relative to the inner binaries. Furthermore, the eccentricities of wide tertiaries are consistent with a thermal eccentricity distribution ($f(e)\propto e$), similar to the typical wide binaries at the same separations. Using this information, below we discuss two particular formation scenarios for close binaries and their wide tertiaries: the dynamical unfolding of compact triples and the Kozai-Lidov mechanism.

\section{Discussion}

\label{sec:discussion}

\subsection{Dynamical unfolding of compact triples}

When three stars are born in a compact configuration, they undergo a chaotic dynamic evolution, sometimes ejecting one star from the rest of the binary system. If the star is not completely ejected, it may stay at larger separations ($\sim10^3$\,AU) and form a wide tertiary around close binaries \citep{Reipurth2012}. If this scenario is a dominant formation channel, it may explain the connection between the close binaries with $P_{in}<10$\,days and their wide tertiaries seen in Fig.~\ref{fig:WCF-P}. Furthermore, this scenario is hypothesized as one of the potential causes for the similar metallicity dependence of the close and wide binary fractions at [Fe/H]$>0$ \citep{Hwang2021a}.

The dynamical unfolding of compact triples predicts highly eccentric wide tertiaries at $>10^3$\,AU, and most of them would have $e>0.8$ at $10^4$\,AU \citep{Reipurth2012}. The wide tertiaries might be randomly oriented relative to the inner binaries due to the chaotic nature of the three-body interaction, consistent with our findings. However, we do not find a significant difference in the eccentricity distribution of the wide tertiaries around close binaries compared to typical wide binaries. Therefore, our results suggest that the dynamical unfolding of compact triples is not the dominant channel for forming close binaries with $P_{in}<10$\,days and their wide tertiaries.

The dynamical unfolding is also difficult to explain the enhanced occurrence rate of double-close-binary quadruples (2+2 systems). For example, the occurrence rate of resolved double-eclipsing-binary quadruples, where two inner eclipsing binaries (with periods $<1$\,days) form a $10^3$-$10^4$\,AU comoving pair, is a factor of $7.3\pm2.6$ higher than the expected value from random pairing \citep{Hwang2020c,Fezenko2022}. The age dependence of the short-period binaries \citep{Hwang2020a} can explain the enhancement for a factor of 2, and the rest enhancement of a factor of $\sim3$ can be explained by the result here that the close binary fraction is $2.33\pm0.11$ times higher among wide pairs than the field close binary fraction. 2+2 systems are found in other parameter space \citep{Cagas2012,Borkovits2018,Powell2021,Kostov2022} and statistical analysis also reports the excess of 2+2 systems \citep{Tokovinin2014b}. Therefore, even if the dynamical unfolding may work for initially compact three-body systems to form hierarchical triples, it is seemingly challenging to have initially compact four-body systems unfold to the 2+2 architecture dynamically.

To summarize, based on the eccentricity of wide tertiaries and the occurrence rate of 2+2 systems, we conclude that the enhanced wide tertiary fraction around close binaries at $P_{in}<10$\,days is not due to the dynamical unfolding of compact triples. One implication is that, after this scenario is ruled out, one remaining hypothesis for the relation between metallicity and wide binary fraction at [Fe/H]$>0$ \citep{Hwang2021a} is radial migration of Galactic orbits \citep{Sellwood2002}. In particular, recent studies show that stellar ages are youngest in solar-metallicity stars and are older toward both low- and high-metallicity ends in the solar neighborhood \citep{Feuillet2018,Xiang2022}, likely due to the radial migration such that the most metal-rich stars are from the inner Milky Way. This age-metallicity relation mimics the metallicity-wide binary fraction relation where the wide binary fraction peaks at the solar metallicity and decreases toward both low- and high-metallicity ends \citep{Hwang2021a,Hwang2022halo}, suggesting a possible common origin.

\subsection{The Kozai-Lidov mechanism}

In a hierarchical triple, the eccentricity of the inner binary may be excited due to the perturbation of the tertiary companion, the so-called Kozai-Lidov mechanism \citep{Kozai1962, Lidov1962}. When the inner binary reaches a high eccentricity so that the pericenter distance is only a few stellar radii, then the tidal friction can circularize the orbit, thus shrinking the orbit and forming a close binary \citep{Harrington1968,Kiseleva1998, Eggleton2001,Fabrycky2007}. The Kozai-Lidov mechanism can also explain the formation of double-close-binary (2+2) systems, where both close binaries serve as the tertiary companion of each other and undergo mutual Kozai-Lidov cycles \citep{Fang2018}.

There are a few conditions for the Kozai-Lidov mechanism to be effective. First, the initial mutual inclination between the inner binary and the outer tertiary needs to be in the range $40^\circ-140^\circ$. A co-planar triple would not undergo the Kozai-Lidov mechanism. Second, the Kozai-Lidov oscillation timescale has to be shorter than the timescale of relativistic pericenter precession, which depends on the separations of inner binaries and tertiaries \citep{Fabrycky2007}. The classical Kozai-Lidov mechanism is a quadrupole effect and does not depend on the eccentricity of the tertiary. The octupole effect, the so-called eccentric Kozai-Lidov effect, is present when the outer tertiary has non-zero eccentricity, and it can lead to chaotic dynamics like orbital flipping \citep{Ford2000,Naoz2011,Naoz2016}.

For close binaries formed from the Kozai-Lidov mechanism with tidal friction, the mutual inclination between tertiaries and inner binaries has enhanced peaks at $\sim40^\circ$ and $\sim140^\circ$, and the peaks are stronger when the initial separations of the inner binaries are larger \citep{Fabrycky2007}. Therefore, the peaks in the mutual inclination distribution serve as an important prediction from the Kozai-Lidov mechanism, although such peaks become less prominent when the octupole effect is taken into account \citep{Naoz2012,Naoz2014}.

\begin{figure}
    \centering
    \includegraphics[width=\linewidth]{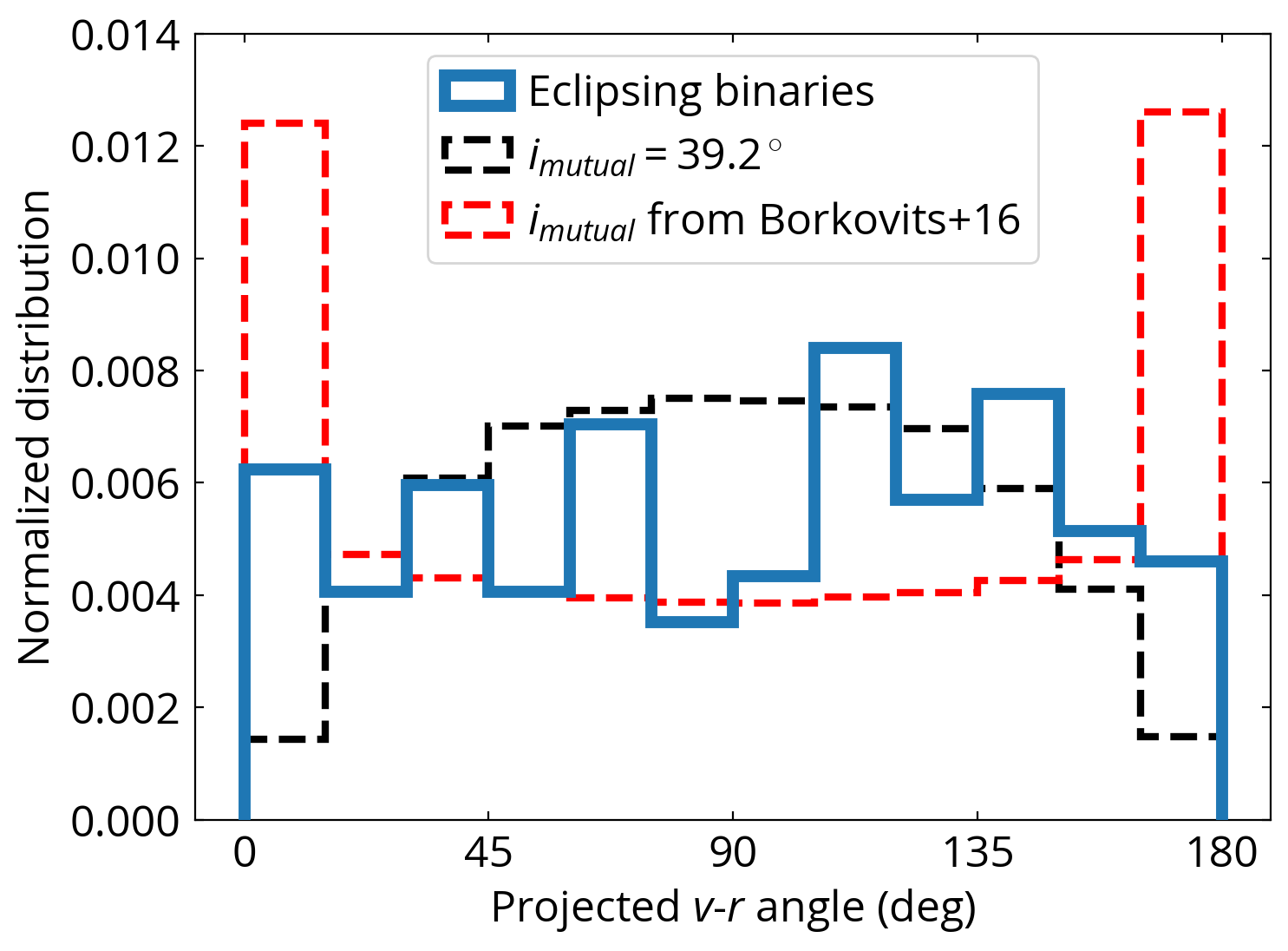}
    \caption{Comparison between the observed $v$-$r$ angle distribution of the wide tertiaries around Gaia eclipsing binaries (solid blue) and the simulated $v$-$r$ distributions (dashed black and dashed red) for different mutual inclinations between the outer companions and the inner eclipsing binaries. The dashed black line considers single-valued mutual inclination of $i_{mutual}=39.2^\circ$ and the dashed red line considers the observed mutual inclination distribution from eclipse timing variation \citep{Borkovits2016}. }
    \label{fig:vr-sim}
\end{figure}

Fig.~\ref{fig:vr-sim} shows the simulated projected $v$-$r$ angle distribution of wide tertiaries with different mutual inclinations relative to the inner edge-on eclipsing binaries. The simulation considers random tertiary orientation (except for the given mutual inclinations) and samples the orbital phase uniformly in time, with an assumed thermal eccentricity distribution for tertiaries. The code is available on GitHub \footnote{https://github.com/HC-Hwang/Eccentricity-of-wide-binaries}.

The black dashed line in Fig.~\ref{fig:vr-sim} represents the case where all tertiaries have single-valued mutual inclinations $i_m=39.2^\circ$, which is the critical angle in the Kozai-Lidov mechanism. The $v$-$r$ angle distribution is relatively flat between $45^\circ$ and $135^\circ$, and has some deficit below $45^\circ$ and above $135^\circ$ due to the lack of co-planar systems. The red dashed line shows the $v$-$r$ angle distribution where the mutual inclinations are sampled from the eclipse timing measurements of \cite{Borkovits2016}. Due to the high fraction of co-planar triples ($\sim50$\% of their triples have $i_m<15^\circ$) in their sample, its $v$-$r$ angle distribution strongly peaks around $0^\circ$ and $180^\circ$. The peaks close to $0^\circ$ and $180^\circ$ are the consequence of edge-on tertiary orbits and are weakly dependent on the assumed eccentricity distribution. The tertiaries in \cite{Borkovits2016} have orbital periods $\sim 2000$\,days (separations $\sim4$\,AU), thus probing a very different tertiary separation range than our wide tertiaries. The eccentricity distribution of their tertiaries is flatter with a peak at $e\sim0.3$, also different from the thermal eccentricity distribution for tertiaries at $\sim10^3$\,AU (Fig.~\ref{fig:alpha}). The K-S test suggests that the observed $v$-$r$ angle distribution is significantly different from the simulated distribution based on \cite{Borkovits2016} (red dashed line), with a $p$-value $<10^{-7}$. The difference between the observed $v$-$r$ distribution and the one from $i_m=39.2^\circ$ (black dashed line) is marginally significant, with a $p$-value of 0.019 from the K-S test.

The lack of the peaks around $0^\circ$ and $180^\circ$ in the observed $v$-$r$ angle distribution (blue) in Fig.~\ref{fig:vr-sim} suggests that the orbits of the wide tertiaries do not preferentially align with their inner eclipsing binaries. Furthermore, as discussed in Sec.~\ref{sec:vr}, the wide tertiaries are consistent with isotropic orientations relative to inner binaries. Therefore, these results suggest that the orientations of tertiaries are more co-planar at tertiary separations of a few AU \citep{Borkovits2016}, and change to isotropic at $\sim10^3$\,AU.

The difference between the observed (solid blue) and the simulated distribution (dashed black) can be explained by that (1) the $i_m$ distribution in reality is not single-valued; or (2) the actual eccentricity distribution is different from the assumed thermal eccentricity distribution. Furthermore, the Kozai-Lidov mechanism may not necessarily cause a strong peak in the mutual angles when the octupole effect is present \citep{Naoz2012,Naoz2014}. Therefore, we cannot draw a strong conclusion about whether the observed $v$-$r$ angle distribution agrees or disagrees with the Kozai-Lidov mechanism.

\cite{Naoz2014} show that the eccentric Kozai-Lidov mechanism does not significantly change the eccentricity distribution of tertiaries, consistent with our findings that the wide tertiaries around these close binaries are similar to typical wide binaries at similar separations (Fig.~\ref{fig:alpha}).  Furthermore, our results show that wide tertiaries around close binaries often have sufficiently high eccentricities where the octupole effect is dynamically important.

For the Kozai-Lidov mechanism to be effective for wide tertiaries at $a_{out}>10^3$\,AU, the most challenging part is its long Kozai-Lidov timescale. To have the Kozai-Lidov cycle timescale shorter than the relativistic precession timescale, the initial inner binary separation needs to be $a_{in,init}>3$\,AU for $a_{out}=10^3$\,AU, or $a_{in,init}>17$\,AU for $a_{out}=10^4$\,AU \citep{Fabrycky2007, Naoz2016}, assuming all component stars have 1\Msun\ and $e=0.5$ for both inner and outer orbits (note that $a_{out}$ does not change during the Kozai-Lidov cycle). In these cases, the Kozai-Lidov timescale is at most $3$\,Gyr, sufficient for stars with typical ages of several Gyr to go through a few Kozai-Lidov cycles to shrink the orbit by the tidal friction. Furthermore, observationally there is a lack of young close ($P_{in}<1$\,days) binaries at ages $\lesssim1$\,Gyr \citep{Hwang2020a}, which may be the orbital migration timescale of either magnetic braking \citep{Stepien1995} or the $\sim$Gyr Kozai-Lidov timescale.

To have a pericenter distance two orders of magnitude smaller than the initial separations $a_{in,init}>3$\,AU so that the tidal effect can be effective, the maximum eccentricity during the Kozai-Lidov cycle needs to be $e_{max}>0.99$. For an isotropic tertiary orientation, only $\sim 2(1-e_{max})<2$\% of triple systems would reach this $e_{max}$ \citep{Munoz2016}. We can estimate how many initial triple systems are needed to explain the observed number of short-period binaries with wide tertiaries. The eclipsing binary fraction (including less edge-on ellipsoidal variables) at $P_{in}<1$\,days is $\sim0.4$\% \citep{Kirk2016}, and $\sim10$\% of eclipsing binaries (thus 0.04\% of the field stars) have wide tertiaries at $>10^3$\,AU (Fig.~\ref{fig:WCF-P}). Since only $<2$\% of the triples can reach $e_{max}>0.99$, we need a $>0.04\%\times50=2$\%\ triple fraction (fraction of triple systems among the field stars) with sufficiently large $a_{in,init}$. The octupole Kozai-Lidov effect can further shorten the Kozai-Lidov timescale and boost the efficiency of orbital migration \citep{Naoz2014}. This 2\% triple fraction seems reasonable compared to the observed triple fraction of $8\pm1$\% in solar-type stars \citep{Raghavan2010,Tokovinin2014b}. Therefore, this simple estimate suggests that the observed triple fraction may be sufficient to form the close binaries at $P_{in}<1$\,days with wide tertiaries through the Kozai-Lidov mechanism where $a_{in,init}$ is sufficiently large.

One difficulty for the Kozai-Lidov scenario to explain is that the wide tertiary excess is relatively flat with respect to tertiary separations (Fig.~\ref{fig:WCF-sep}). Even more puzzling, the wide tertiary excess of eclipsing binaries seems to peak around $10^4$\,AU, although a larger sample is needed to confirm the signal. In the Kozai-Lidov mechanism, the tertiary excess is expected to be higher at smaller tertiary separations where the Kozai-Lidov timescale is shorter and thus a wider range of $a_{in, init}$ can excite the Kozai-Lidov oscillation.

Another challenge for the Kozai-Lidov scenario is the reduced wide tertiary fraction at $P_{in}\sim10^{2.5}$\,days (Fig.~\ref{fig:WCF-P}). One possible explanation is that close binaries with wide tertiaries at initial $P_{in,init}\sim10^{2.5}$\,days have migrated to shorter orbital periods, leaving close binaries currently at $P_{in}\sim10^{2.5}$\,days having fewer tertiaries. At $P_{in,init}\sim10^{2.5}$\,days ($a_{in,init}\sim1$\,AU), the quadrupole Kozai-Lidov timescale is longer than the relativistic precession timescale, and thus the octupole effect with shorter Kozai-Lidov timescales may be necessary. The reduced tertiary fraction is not due to the dynamical stability because even at $P_{in}=10^2$\,days, their separation ratios are $a_{out}/a_{in}\gtrsim1000$, significantly above the three-body stability criterion \citep{Mardling2001, Hayashi2022}. The literature simulations for the Kozai-Lidov mechanism have not well explored the parameter space at large $a_{in,init}$ and large $a_{out}$, and future work is needed to investigate if the different initial conditions can explain these results.

Here we mainly discuss the Kozai-Lidov mechanism in the three-body system, but some of the triples investigated here may be higher-order multiples consisting of more than three bodies. It is possible that the presence of wide companions at $>10^3$\,AU is correlated with the occurrence of another object at $\lesssim 10^2$\,AU, and this object is responsible for driving the Kozai-Lidov cycle of the inner binaries, resulting in the enhanced wide tertiary fraction in close binaries at $P_{in}<10$\,days. There is another statistical effect that makes the wide companion fraction higher if the sample preferentially avoids close companions. For example, hot Jupiter hosts have a higher (wide) binary fraction \citep{Ngo2016, Belokurov2020,Hwang2020c}, which may be due to that hot Jupiters cannot form in close binaries \citep{Moe2021binaryplanet}. \cite{Moe2021binaryplanet} further point out that this statistical effect is small for close binaries because they do not have a strong deficit of nearby companions. Thus the excess of wide tertiaries around close binaries is physical.

\section{Conclusions}
\label{sec:conclusion}

In this paper, we use close binaries from Gaia DR3 to investigate the mysterious interplay with their wide tertiaries at $\gtrsim 10^3$\,AU. In particular, we investigate the wide tertiary fraction across four orders of magnitude in inner binary periods. Then we use the $v$-$r$ angles of the wide tertiaries to constrain their orientations and eccentricities. Our findings are as follows:

\begin{enumerate}
    \item The wide ($10^3$-$10^4$\,AU) tertiary fraction increases with decreasing orbital periods of the inner binaries (Fig.~\ref{fig:WCF-P}). The wide tertiary fraction of eclipsing binaries (median $P_{in}=0.41$\,day) is $2.33\pm 0.11$ times higher than the field wide binary fraction. The separation distributions of the wide tertiaries are similar to that of wide binaries, except that eclipsing binaries may have an excess of tertiaries around $10^4$\,AU (Fig.~\ref{fig:WCF-sep}). 
    
    \item The $v$-$r$ angle distributions of the wide tertiaries are similar among different categories of close binaries, which have different orbital periods and inner binary orientations (Fig.~\ref{fig:vr}). We conclude that the wide tertiaries have relatively random orientation relative to the inner binaries, and the inferred eccentricity distribution is close to thermal, similar to typical wide binaries at similar separations (Fig.~\ref{fig:alpha}).
    \item We consider two scenarios, the dynamical unfolding of compact triples and the Kozai-Lidov mechanism, because they may explain the enhanced wide tertiary fraction around close binaries with $P_{in}<10$\,days. However, the dynamical unfolding scenario is disfavored because the observed wide tertiaries do not have particularly high eccentricities, inconsistent with the expectation from the dynamical unfolding of compact triples. 
    \item For wide tertiaries at $\gtrsim 10^3$\,AU, the initial separations of the inner binaries need to be $>3$\,AU so that the relativistic pericenter precession does not suppress the Kozai-Lidov oscillation. Our estimate suggests that the observed triple fraction in the field stars may be sufficient for this process to form the observed number of close binaries ($P_{in}<1$\,days) with wide tertiaries at $>10^3$\,AU. However, the Kozai-Lidov scenario is challenging to explain the flat tertiary excess with respect to tertiary separations and the tentative enhanced excess at $10^4$\,AU for the tertiaries of eclipsing binaries. Furthermore, the octupole effect may be needed to explain the reduced wide tertiary fraction at $P_{in}\sim10^{2.5}$\,days. Another possibility is that some triple systems investigated in this paper may be higher-order multiples. The presence of wide tertiaries at $>10^3$\,AU can be correlated with another additional companion at $\lesssim10^2$\,AU, and this additional closer companion may be responsible for driving the Kozai-Lidov cycle of the inner binary.
    
\end{enumerate}

\section*{ACKNOWLEDGEMENTS}

The author is grateful to the referee for the constructive report. HCH appreciates the discussions with Chris Hamilton, Nadia Zakamska, Roman Rafikov, Joshua Winn, Scott Tremaine, Sihao Cheng, and Andrei Tokovinin. HCH acknowledges the support of the Infosys Membership at the Institute for Advanced Study. 

\section*{Data Availability}

The data underlying this article are available online. The datasets were derived from sources in the public domain: Gaia Data Archive https://gea.esac.esa.int/archive/

\appendix

\restartappendixnumbering
\setcounter{figure}{0}

\section{Gaia query for nearby sources around close binaries}

\label{sec:appendix-query}

Below is the Gaia query for the nearby sources within 10\,arcsec around eclipsing binaries (\texttt{non\_single\_star = 4}):

{\obeylines\obeyspaces
	\texttt{
		SELECT gaia.source\_id, gaia.ra, gaia.dec, gaia.l, gaia.b, gaia.parallax, gaia.phot\_g\_mean\_mag, nss.non\_single\_star, gaia.pmra, gaia.pmdec, nss.parallax as parallax0, nss.pmra as pmra0, nss.pmdec as pmdec0,
DISTANCE(
    POINT(nss.ra, nss.dec),
    POINT(gaia.ra, gaia.dec)
) * 3600. AS dist\_arcsec
FROM gaiadr3.gaia\_source\_lite AS nss
JOIN gaiadr3.gaia\_source\_lite AS gaia
ON (1 = 
CONTAINS(
    POINT(nss.ra, nss.dec),
    CIRCLE(gaia.ra, gaia.dec, 10 / 3600.)
)) AND
(gaia.source\_id != nss.source\_id)
WHERE nss.non\_single\_star = 4
}
}

\section{Wide tertiaries at other separations}
\label{sec:appendix}

\begin{figure*}
    \centering
    \includegraphics[width=0.49\linewidth]{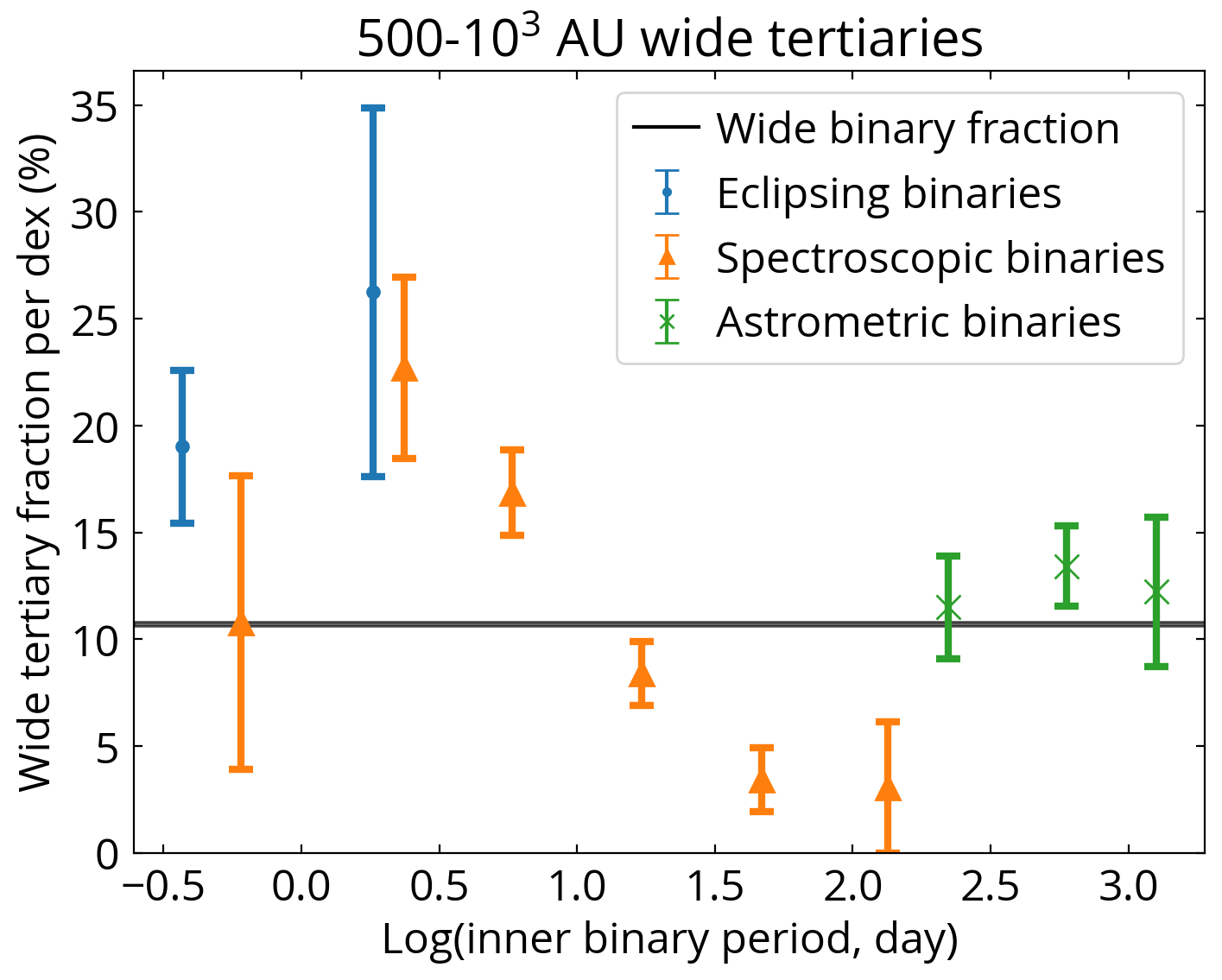}
    \includegraphics[width=0.49\linewidth]{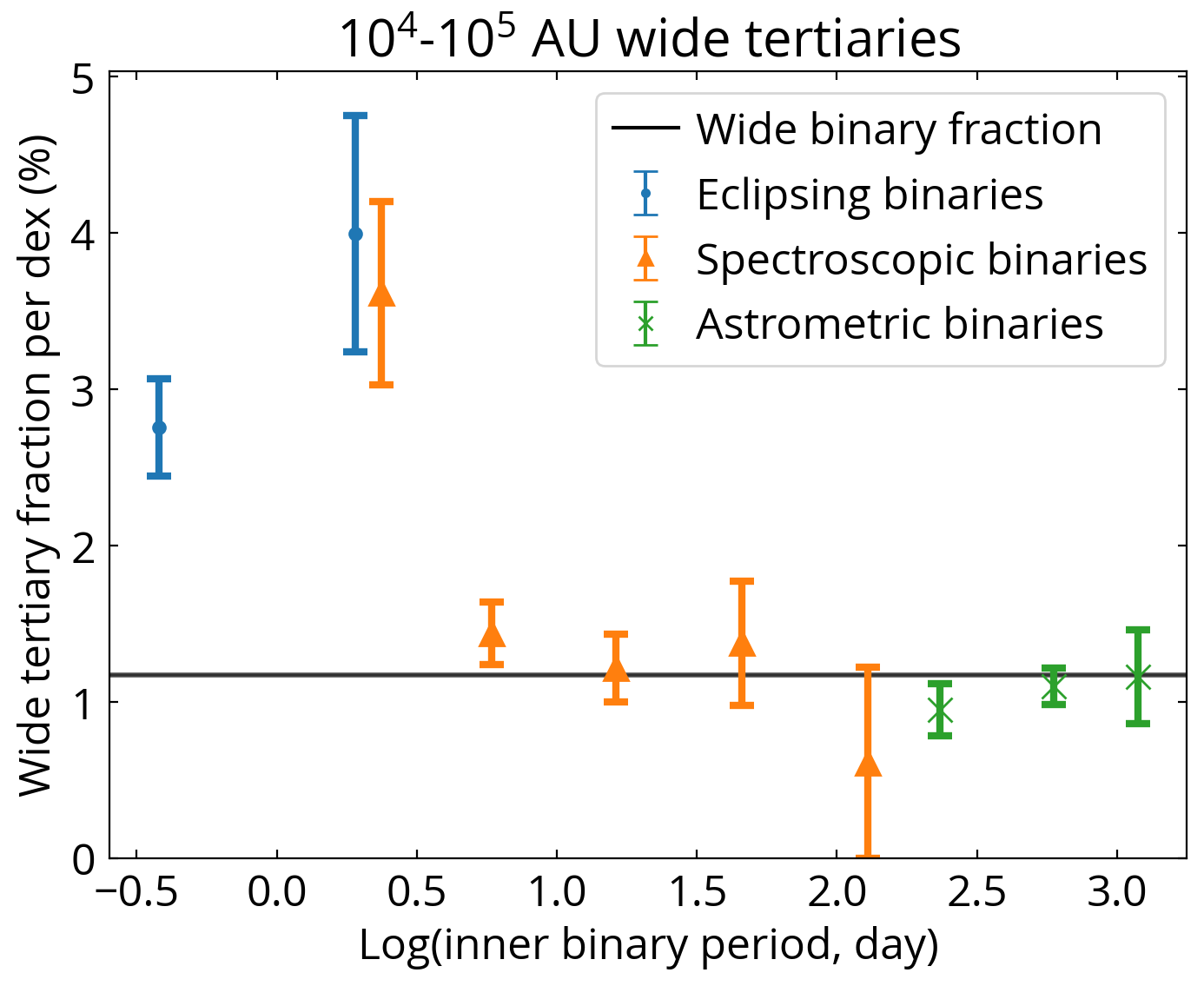}
    \caption{The wide tertiary fractions per tertiary separation dex for tertiaries at 500-$10^3$\,AU (left) and $10^4$-$10^5$\,AU (right). The overall trend is similar to Fig.~\ref{fig:WCF-P}, with the wide tertiary fraction increasing with decreasing inner binaries' periods.  }
    \label{fig:WCF-P-appendix}
\end{figure*}

In the main text, we consider the wide tertiaries at $10^3$-$10^4$\,AU. Here we consider wide tertiaries at other separations. In Fig.~\ref{fig:WCF-P-appendix}, the left panel considers wide tertiaries at 500-$10^3$\,AU and the right panel at $10^4$-$10^5$\,AU. For the left panel, we adopt a parallax cut $>4$\,mas (distances $<250$\,pc) so that the wide tertiaries have angular separations $>2$\,arcsec. The right panel still uses the parallax cut $>2$\,mas (distances $<500$\,pc), and therefore the angular separations of the wide tertiaries are $>20$\,arcsec. Other criteria in Sec.~\ref{sec:sample-close} like the main-sequence cut are still used here. These results are corrected for completeness. Following Sec.~\ref{sec:sample-wide}, we do not apply the \texttt{R\_chance\_align} criterion here. While the contamination from change alignments is non-negligible for wide tertiaries at $>10^4$\,AU, we find that applying \texttt{R\_chance\_align}$<0.1$ (i.e. the contamination rate $<10$\% for every wide tertiary) reduces the wide tertiary fraction by a factor of $\sim1.5$ in the right panel, but the overall trend and the tertiary excess at $P_{in}<10^{0.5}$\,days remain unchanged. We use the same x-axis bins as in Fig.~\ref{fig:WCF-P} and discard points with error bars larger than 10\% in the left panel and 1\% in the right panel.

The overall trend in Fig.~\ref{fig:WCF-P-appendix} is similar to Fig.~\ref{fig:WCF-P}, with the wide tertiary fraction increasing with decreasing inner binaries' period. Similar to Fig.~\ref{fig:WCF-P}, the left panel shows that the 500-$10^3$\,AU wide tertiary fraction reaches the field star level at $P_{in}=10$\,days. Interestingly, the $10^4$-$10^5$\,AU wide tertiary fraction (right panel) is higher than the wide binary fraction only at $P_{in}<10^{0.5}$\,day.

\bibliography{WideAroundCloseBinary}{}
\bibliographystyle{aasjournal}
\end{document}